\documentclass[12pt]{article}
\def\beq{\begin{equation}}
\def\eeq{\end{equation}}
\def\bea{\begin{eqnarray}}
\def\eea{\end{eqnarray}}
\def\bq{\begin{quote}}
\def\eq{\end{quote}}

\def\gappeq{\mathrel{\rlap {\raise.5ex\hbox{$>$}}
{\lower.5ex\hbox{$\sim$}}}}

\def\lappeq{\mathrel{\rlap{\raise.5ex\hbox{$<$}}
{\lower.5ex\hbox{$\sim$}}}}

\begin{document}
\topmargin -0.5cm
\oddsidemargin -0.3cm
\pagestyle{empty}
\begin{flushright}
\end{flushright}
\vspace*{5mm}
\begin{center}
\large
{\bf On the existence of Hamiltonians for non-holonomic systems} \\
\normalsize
\vspace*{1.5cm} 
{\bf Christofer Cronstr\"{o}m}$^{*)}$ \\
\vspace{0.3cm}
Helsinki Institute of Physics\\
P. O. Box 64, FIN-00014 University of Helsinki, Finland \\
\vspace{0.3cm}
and\\
\vspace{0.3cm}
{\bf Tommi Raita}$^{**)}$\\
\vspace{0.3 cm}
Physics Department\\
P. O. Box 64, FIN-00014 University of Helsinki, Finland \\
\vspace{0.3cm}
 
\vspace*{1cm}

{\bf ABSTRACT} \\

\end{center}
We consider the existence of  Hamiltonians for autonomous non-holonomic mechanical systems. The approach is elementary in  that the existence of a Hamiltonian for a  non-holonomic system is equivalent to the existence of an appropriate Lagrangian for the system in question. The existence of such a Lagrangian is related to the inverse problem of constructing a Lagrangian from the equations of motion. A simple example in three dimensions with one non-holonomic constraint is analyzed in detail. In this case there is no Lagrangian reproducing the equations of motion in three dimensions. Thus the system does not admit a variational formulation in three dimensions. However, the system in question is equivalent to a two-dimensional system which does admit a variational formulation. Two distinct  Lagrangians and their corresponding  Hamiltonians are constructed explicitly for this two-dimensional system.
\vspace*{5mm}
\noindent

\vspace*{1cm} 
\noindent 

\noindent 
$^{*)}$ e-mail address: Christofer.Cronstrom@Helsinki.fi \\ 
$^{**)}$ e-mail address: Tommi.Raita@Helsinki.fi

\vfill\eject

\pagestyle{plain}
\pagenumbering{arabic}
\setcounter{page}{1}

\section{Introduction}
Hamilton's principle for mechanical systems with non-holonomic constraints has  recently been discussed by Flannery \cite{Flannery}.  In particular a variational formulation of the equations of motion of a mechanical system was discussed both for holonomic and non-holonomic constraints. It was shown  that while the equations of motion for a system with holonomic constraints can be obtained as variational equations,  with the  constraints being taken into account by the multiplication rule in the calculus of variations  \cite{Mikhlin},  the corresponding procedure with non-holonomic constraints leads to equations which differ  from the correct equations of motion.  

The problems discussed by Flannery are not new;  they have been discussed in the literature at least since Hertz's textbook \cite{Hertz},  in which the use of variational principles in mechanics was questioned. Two papers published  by Jeffreys \cite{Jeffreys} and Pars \cite{Pars} consider Hamilton's principle for non-holonomic systems, and propose  rectification  of previous papers in which the variational procedures discussed by Flannery had been used also  for non-holonomic systems.   

Several papers  have advocated the use of a variational principle involving the multiplication rule in the calculus of variations for non-holonomic systems.  In addition to the papers of this kind quoted by Flannery and by Pars and Jeffreys, respectively,  we  mention a paper by Berezin  \cite{Berezin},  in which  no distinction is made between holonomic and non-holonomic systems. 

It is also appropriate to to mention that,  in contradistinction to the original "Classical Mechanics" by Goldstein  \cite{Goldstein59},  the 3rd edition of this classical mechanics textbook advocates  the use of a variational principle involving the the multiplication rule for non-holonomic systems  \cite{newGoldstein}.  However, the use of this principle for non-holonomic systems was later retracted  \cite{retraction}. This fact was pointed out already by Flannery. 

It appears that if a system with non-holonomic constraints does not admit a variational formulation, then the dynamics of the system is not governed by a Hamiltonian $H$. This is the question we address in this paper: Can a non-holonomic system be described in terms of   Hamiltonian equations of motion?  We confine the detailed discussion to a simple example in three dimensions introduced by Pars  \cite{Pars}.  We show that in this case the equations of motion are reducible to a set of equations for a two-dimensional autonomous system, which \underline{can} be formulated as Hamiltonian equations. However, the original equations of motion in three dimensions do not admit a Hamiltonian formulation.

Our analysis is elementary in that the existence of a Hamiltonian for a given non-holonomic system is considered to be equivalent to the existence of an appropriate Lagrangian $L(q, \dot{q})$ for the system in question. By  appropriate is meant that the Lagrangian is non-degenerate, {\em i.e.} that the equations defining the canonical momenta $p^{j}$, 
\beq
\label{nondeg}
p^{j} := \frac{\partial L}{\partial \dot{q}_{j}},
\eeq
are solvable for the generalized velocities $\dot{q}_{j}$. It should be noted that we discuss only autonomous systems. Hence the Lagrangian is allowed to depend on time only through the coordinates $q$ and velocities $\dot{q}$. 

 The existence of an appropriate Lagrangian is related to the inverse problem of constructing a Lagrangian from the appropriate equations of motion. To the best of our knowledge, a  complete  solution to the inverse problem is not known in the general n-dimensional case for $n \geq 3$. 

In the next section we consider the  Lagrange equations of motion for an autonomous mechanical system with both holonomic and non-holonomic constraints.  This problem was considered  in some detail by Flannery \cite{Flannery}. For the sake of completeness  we  consider the  equations obtained from the generalized form of  d'Alembert's principle and the equations which follow from a variational procedure with constraints implemented by the multiplication rule. In the non-holonomic case these equations are not identical. 

\section{Lagrange equations with  constraints}

Consider an autonomous  mechanical system with independent generalized coordinates $q_{1},...,q_{n}$, and  velocities $\dot{q}_{1},...,\dot{q}_{n}$. Let the kinetic energy  be  $T$, and the generalized applied forces on the system be  $Q^{j}, j = 1,...,n$. The generalized principle of d'Alembert (see  e.g.  the classical texts by Goldstein \cite{Goldstein59} or Whittaker \cite{Whittaker}) then gives the following equation,
\beq
\label{Lag0}
\sum_{j=1}^{n} \left \{\frac{d}{dt}\left (\frac{\partial T}{\partial \dot{q}_{j}} \right ) - \frac{\partial T}{\partial q_{j}} - Q^{j}\right \} \delta q_{j} = 0,
\eeq
where the quantities $\delta q_{j}$ are virtual displacements of the system. If the virtual displacements $\delta q_{j}, j = 1,\ldots, n$ are independent, then the equation (\ref{Lag0}) results in the Lagrange equations of motion,
\beq
\label{Lag1}
\frac{d}{dt}\left (\frac{\partial T}{\partial \dot{q}_{j}} \right ) - \frac{\partial T}{\partial q_{j}} = Q^{j}, \; j = 1, \ldots, n,
\eeq

We  generalize  to systems with $1 \leq m < n$ independent non-holonomic constraints, which are taken to be linear and homogeneous in the velocities. The constraint equations are of the following form,
\beq
\label{nholco}
\sum_{j=1}^{n} a_{i}^{~j}(q_{1},...,q_{n}) \dot{q}_{j} = 0,\,  i = 1,..., m < n,
\eeq
where the quantities $a_{i}^{~j}, i = 1,\ldots,m, j = 1,\ldots,n$, are given functions of the variables $q_{1},...,q_{n}$.

The derivation given below of the equations of motion for this non-holonomic system can be found {\em e.g.} in the textbook by  Whittaker  \cite{Whittaker}.

Implement the constraints (\ref{nholco}) by regarding the system to be acted on by external applied forces  $Q^{j}$ and  by certain additional forces of constraint $Q'^{j},\, j=1,\ldots,n$, which force the system to satisfy the non-holonomic conditions (\ref{nholco}). The equation (\ref{Lag0}) is then replaced by the following equation,
\beq
\label{Lag3}
\sum_{j=1}^{n} \left \{\frac{d}{dt}\left (\frac{\partial T}{\partial \dot{q}_{j}} \right ) - \frac{\partial T}{\partial q_{j}} - Q^{j} - Q'^{j} \right \} \delta q_{j} = 0,
\eeq
In Eq. (\ref{Lag3})  the virtual displacements $\delta q_{j}, j = 1, \ldots, n$,  can now be regarded as independent.  Thus one obtains  the equations of motion,
\beq
\label{Lagnh}
\frac{d}{dt}\left (\frac{\partial T}{\partial \dot{q}_{j}} \right ) - \frac{\partial T}{\partial q_{j}} =Q^{j} + Q'^{j}, \,j = 1, \ldots, n.
\eeq
The forces of constraint, $Q'^{j}, j = 1, \ldots, n$, are {\em a priori} unknown, but they are such that, in any instantaneous displacement $\delta q^{j}, j = 1,\ldots, n$, consistent with the constraints (\ref{nholco}), they do no work.  The non-holonomic constraints (\ref{nholco})  imply the following conditions on the possible instantaneous displacements   $\delta q^{j}, j = 1,\ldots, n$ of the system,
\beq
\label{cdelta}
\sum_{j=1}^{n}a_{i}^{~j}(q_{1},...,q_{n}) \delta q_{j} = 0,\,  i = 1,..., m < n.
\eeq
For any instantaneous displacements $\delta q^{j}, j = 1,\ldots, n$, which satisfy the conditions (\ref{cdelta}), the work $\delta W'$ done by the constraint forces $Q'^{j}, j=1,\ldots,n$ equals zero,
\beq
\label{workc}
\delta W' \equiv \sum_{j=1}^{n} Q'^{j}\delta q^{j} = 0.
\eeq
The conditions (\ref{cdelta}) and (\ref{workc})  together imply that
\beq
\label{cforce}
Q'^{j} = \sum_{i=1}^{m} \lambda^{i} a_{i}^{~j},\, j= 1,\ldots, n,
\eeq
where the quantities $\lambda^{i}, i = 1,\ldots, m$, are time-dependent parameters.  The equations (\ref{Lagnh}) have  been reduced to
\beq
\label{Lagnh2}
\frac{d}{dt}\left (\frac{\partial T}{\partial \dot{q}_{j}} \right ) - \frac{\partial T}{\partial q_{j}} =Q^{j} + 
\sum_{i=1}^{m} \lambda^{i} a_{i}^{~j}, \,j = 1, \ldots, n.
\eeq
To these $n$  equations of motion one should add the $m$ equations of constraint (\ref{nholco}). We have $n+m$ equations for the determination of $n+m$ quantities $q_{j}(t), j = 1,\ldots,n$, and $\lambda^{i}(t), i = 1, \ldots,m$. 

It should be observed that in the argument above, one has not required the constraint equations (\ref{nholco}) to be in force under general  variations $q_{j} \rightarrow q_{j} + \delta q_{j}$; the constraints (\ref{nholco}) are only imposed on the actual motion of the system. 

Now assume that the external applied forces $Q^{j}, j=1,\ldots,n$,  can be expressed in terms of a potential $V$ such that,
\beq
\label{GenV}
Q^{j} = - \frac{\partial V}{\partial q_{j}} +  \frac{d}{dt}\left (\frac{\partial V}{\partial \dot{q}_{j}}\right ), j = 1,...,n.
\eeq
Using the notation
\beq
\label{ell0}
L_{0} := T - V,
\eeq
the equations  (\ref{Lagnh2}) can be written as,
\beq
\label{Lagnh3}
\frac{d}{dt}\left (\frac{\partial L_{0}}{\partial \dot{q}_{j}}\right ) - \frac{\partial L_{0}}{\partial q_{j}} = \sum_{i=1}^{m} \lambda^{i} a_{i}^{~j}, \,j = 1, \ldots, n.
\eeq

It should  be observed that the $m$ one-forms (\ref{cdelta})  are  non-integrable by assumption, for otherwise the system would be holonomic. In the integrable case (after multiplying the conditions (\ref{nholco}) with integrating factors if necessary) one would have
\beq
\label{holo}
a_{i}^{~j} = \frac{\partial G_{i}}{\partial q_{j}}, i = 1,\ldots,m,
\eeq
where the functions $G_{i}, ,i=1,\ldots,m$, are $m$ independent functions of the variables $q_{j}, j = 1,\ldots,n$,
\beq
\label{holoc}
G_{i} = G_{i}(q_{1}, \ldots, q_{n}), i = 1, \ldots, m.
\eeq
The $m$ constraint equations (\ref{nholco}) would then be equivalent to the following $m$ holonomic constraints,
\beq
\label{holoc2}
G_{i}(q_{1}, \ldots, q_{n}) = C_{i}, i = 1, \ldots, m,
\eeq
where the quantities $C_{i}, i = 1,\ldots, m$ are constants. In this case  the equations (\ref{Lagnh3}) are the Euler-Lagrange equations of the variational problem
\beq
\label{varco}
\delta \int \,dt L_{0} = 0.
\eeq
under the constraints (\ref{holoc2}). These constraints can be implemented with the multiplication rule in the calculus of variations.  This leads to the following  {\em free}  variational problem with Lagrange multipliers $\lambda^{i}, i =  1,\ldots, m$,
\beq
\label{frvar}
\delta \int \, dt \left [ L_{0} + \sum_{i=1}^{m} \lambda^{i}\left (G_{i} - C_{i} \right ) \right ] = 0.
\eeq
The variational problem yields
\beq
\label{Lagnh5}
\frac{d}{dt}\left (\frac{\partial L_{0}}{\partial \dot{q}_{j}}\right ) - \frac{\partial L_{0}}{\partial q_{j}} - \sum_{i=1}^{m} \lambda^{i} \, \frac{\partial G_{i}}{\partial q_{j}} = 0, \,j = 1, \ldots, n.
\eeq
The system of equations (\ref{Lagnh5}), together with the constraints (\ref{holoc2}),  are  the correct equations of motion for the system under consideration in the integrable (holonomic) case. These equations  are a set of Euler-Lagrange equations with the integrand in Eq. (\ref{frvar}) as a Lagrangian $L$,
\beq
\label{Lagrangh}
L := L_{0} +   \sum_{i=1}^{m} \lambda^{i}\left (G_{i} - C_{i} \right ),
\eeq
provided one adjoins the time-dependent parameters $\lambda^{i}, i = 1,\ldots, m$, as new coordinates to the system.  It should  be noted that the usual method of transition to a Hamiltonian from the Lagrangian (\ref{Lagrangh}) does not apply, since the momenta conjugate to the new coordinates $\lambda^{j}$ vanish identically.

Contrary to the assertions in some of the papers referred to in the references  \cite{Flannery}, 
\cite{Jeffreys},  \cite{Pars}, as well as in reference \cite{Berezin},  a similar procedure in the non-holonomic case does not  lead to the correct equations of motion.  Specifically, if one considers the variational problem (\ref{varco}) under the constraints (\ref{nholco}) using the multiplication rule, one is led  to the following free variational problem,
\beq  
\label{varnh3}
\delta \int \,dt \left [L_{0} - \sum_{i=1}^{m}  \mu^{i} \sum_{j=1}^{n}a_{i}^{~j}(q_{1},...,q_{n}) \dot{q}_{j} \right ] = 0,
\eeq
where the Lagrange multipliers are now  denoted by $ \mu^{i}, i = 1,\ldots,m$. The variational equations following from Eq. (\ref{varnh3}) are,
\beq
\label{CC1}
\frac{d}{dt}\left (\frac{\partial L_{0}}{\partial \dot{q}_{j}}\right ) - \frac{\partial L_{0}}{\partial q_{j}} -
\sum_{i=1}^{m} \dot{ \mu}^{i} a_{i}^{~j} 
- \sum_{i=1}^{m}  \mu^{i}\sum_{k=1}^{n}\left ( \frac{\partial a_{i}^{~j}}{\partial q_{k}} - 
\frac{\partial a_{i}^{~k}}{\partial q_{j}} \right ) \dot{q}_{k} = 0.
\eeq
The equations (\ref{CC1}) are not identical  to the correct equations of motion (\ref{Lagnh3}) for the non-holonomic system under consideration. However, if the integrability conditions
\beq
\label{intcond}
\frac{\partial a_{i}^{~j}}{\partial q_{k}} -  \frac{\partial a_{i}^{~k}}{\partial q_{j}} = 0, i = 1, \ldots, m, j, k = 1, \ldots, n,
\eeq
are valid, in which case the system becomes holonomic, the equations of motion (\ref{CC1}) coincide with the corresponding correct equations of motion  (\ref{Lagnh3}) [equivalently Eqns, (\ref{Lagnh5})] upon a change of notation $\dot{ \mu}^{i} \rightarrow   \lambda^{i},  i = 1, \ldots, m$. 

The generalized principle of d'Alembert  differs from the variational principle involving the multiplication rule in the case of non-holonomic constraints.  One consequence of this difference is the fact that the equations of motion following from the principle of d'Alembert differ in form from the equations of motion which follow from the variational principle.   It is not  excluded that these equations may have the same  solutions, however. It thus remains to consider the question of whether the equations of motion (\ref{Lagnh3}) and the variational equations (\ref{CC1}) can have coinciding solutions in general.  In his discussion of this problem  Pars \cite{Pars} used a simple yet non-trivial example in three-dimensional configuration space to show that the equations (\ref{Lagnh3}) and  (\ref{CC1})  in that case can not have coincident general solutions.  We will give a detailed discussion of Pars'  example  below, adding a few details related to the relevance of the initial values. For clarity, we also pay attention to the dimensions of the quantities in the example, by including  appropriate dimensional constants. 

\section{Pars' example}

The example considered by Pars \cite{Pars} is the case of an otherwise free particle of  mass $m$ in three-dimensional Euclidean space with coordinates designated by $(x, y, z)$, except that the motion of the particle is subjected to the following non-holonomic constraint,
\beq
z\dot{x} - \ell \dot{y} = 0,
\label{dal70}
\eeq
where $\ell$ is a constant with the dimension length. The dimensional parameters $m$ and $\ell$, which were absent in the formulation given by Pars, are introduced here for clarity.

The Lagrangian $L_{0}$ in this case is the following,
\beq
L_{0} = \frac{m}{2}\left (\dot{x}^{2} + \dot{y}^{2} + \dot{z}^{2} \right ).
\label{dal69}
\eeq
The Lagrange equations of motion (\ref{Lagnh3})  reduce now to the following equations,
\beq
m\ddot{x} = \lambda z\;, m\ddot{y} = - \lambda \ell \;, m\ddot{z} = 0,
\label{dal71}
\eeq  
where $\lambda$ is a parameter, which is  to be determined by solving the system of equations (\ref{dal71}) and (\ref{dal70}). The initial conditions are as follows, 
\beq
(x, y, z)|_{t=0} = 0,
\label{dal72}
\eeq
and
\beq
\label{dal72vel}
(\dot{x}, \dot{y}, \dot{z})|_{t=0} = (u, 0, w),
\eeq
where $u$ and $w$ are parameters at our disposal, except for the conditions $u \neq 0$ and  
$w \neq 0$. If $u=0$ or $w=0$ then the solutions to the equations (\ref{dal71}) with the initial conditions (\ref{dal72}) and (\ref{dal72vel}) are trivial and uninteresting.  It should also be noted that the condition $\dot{y}(0) = 0$ in ({\ref{dal72vel}) above is not  a free choice, but a consequence of the constraint equation (\ref{dal70}) and the initial values (\ref{dal72}).

We now consider an alternative form of the Lagrange equations (\ref{dal71}) and the constraints (\ref{dal70}). Differentiating the constraint equation  (\ref{dal70}) one obtains
\beq
\label{dercons}
z\ddot{x} + \dot{x}\dot{z} - \ell \ddot{y} = 0.
\eeq
Eliminating the quantities $\ddot{x}$ and $\ddot{y}$ from the equation (\ref{dercons}) above with the aid of the equations (\ref{dal71}),  one obtains the following expression for the quantity $\lambda$:
\beq
\label{Mu}
\lambda = - \frac{m \dot{x} \dot{z}}{\ell^{2}+z^{2}}.
\eeq
Inserting  the expression (\ref{Mu}) for the parameter $\lambda$ into the original equations (\ref{dal71})
one obtains the following three equations,
\beq
\label{Alteqn1}
\ddot{x} =   - \frac{\dot{x} z  \dot{z}}{\ell^{2}+z^{2}}, 
\eeq
\beq
\label{Alteqn2}
\ddot{y} =  \frac{\ell \dot{x}\dot{z}}{\ell^{2}+z^{2}}, 
\eeq
and
\beq
\label{Alteqn3}
\ddot{z}  =  0.
\eeq
It should be noted that the mass $m$ does not appear in  the equations (\ref{Alteqn1})  - (\ref{Alteqn3}) .

Before proceeding further, we demonstrate that the set of equations (\ref{Alteqn1})  - (\ref{Alteqn3}) above are essentially equivalent to the original equations (\ref{dal71}) and the constraints (\ref{dal70}). Multiplying the equation  (\ref{Alteqn1}) with $z$, and subtracting the  equation (\ref{Alteqn2})  from the result, one obtains
\beq
\label{integ}
z\ddot{x} - \ell  \ddot{y} = - \dot{x} \dot{z} \Leftrightarrow \frac{d}{dt}(z\dot{x} - \ell \dot{y}) = 0 \Leftrightarrow (z\dot{x} - \ell \dot{y})  = {\cal C},
\eeq
where ${\cal C}$ is a constant. Using finally the initial conditions (\ref{dal72}) and (\ref{dal72vel}) to evaluate this constant one obtains
\beq
\label{Constant}
{\cal C} = 0.
\eeq
The equations (\ref{Alteqn1}) - (\ref{Alteqn3}) thus imply the constraints (\ref{dal70}) when one also uses the information encoded in the initial conditions (\ref{dal72}) and (\ref{dal72vel}). The equations (\ref{Alteqn1}) - (\ref{Alteqn3}) are indeed of the form (\ref{dal71}), where the parameter $\lambda$ is identified with the expression (\ref{Mu}). We have now demonstrated that the equations (\ref{Alteqn1}) - (\ref{Alteqn3}) together with the initial conditions  (\ref{dal72}) and (\ref{dal72vel}) are equivalent to the original equations (\ref{dal71}) and the constraints (\ref{dal70}). The wording "essentially equivalent" used above was meant to reflect  the fact that one had to invoke the initial conditions (\ref{dal72}) and (\ref{dal72vel}) in order to show that the constraints (\ref{dal70}) are a consequence of the alternative equations (\ref{Alteqn1}) - (\ref{Alteqn3})  and not a separate condition, as in the formulation (\ref{dal71}) which involves the parameter $\lambda$.

We note that the  equation (\ref{Alteqn1}) can be integrated, yielding
\beq
\label{xdot}
\dot{x}\sqrt{\ell^{2}+z^{2}} = \ell u,
\eeq
were $u$ is the  initial value at $t=0$ for the quantity  $\dot{x}$.  The result (\ref{xdot}) will be used shortly.

Consider then the variational problem (\ref{varnh3}) for the case at hand, {\em i.e.}
\beq
\label{CC1var}
\delta \int  dt \left [L_{0} - \mu (z\dot{x} - \ell \dot{y}) \right ] = 0,
\eeq
where the function $L_{0}$ is given in Eq. (\ref{dal69}). The differential equations which follow from 
Eq. (\ref{CC1var}) are the following
\beq
\label{mueq1}
\frac{d}{dt}\left ( m\dot{x} - \mu z \right ) =  0, 
\eeq
\beq
\label{mueq2}
\frac{d}{dt} \left ( m \dot{y} + \mu \ell \right )  =  0,
\eeq
and
\beq
\label{mueq3}
m\ddot{z}  + \mu \dot{x}  =  0.
\eeq
To the equations (\ref{mueq1}) - (\ref{mueq3}) one should still add the constraint equation (\ref{dal70}).

It will be shown that the solutions to the equations (\ref{Alteqn1}) - (\ref{Alteqn3}) with the initial conditions (\ref{dal72}) and (\ref{dal72vel}) can not satisfy the variational equations (\ref{mueq1}) - 
(\ref{mueq3}) and  the constraint equation (\ref{dal70}), except in certain trivial cases cases, as shown below. 

Assume now that there are appropriate  solutions $x(t)$, $y(t)$ and $z(t)$, which satisfy both  sets of equations  (\ref{Alteqn1}) - (\ref{Alteqn3}) and (\ref{mueq1}) - (\ref{mueq3}) together with the constraint   (\ref{dal70}), respectively, under the initial conditions  (\ref{dal72}) and (\ref{dal72vel}). From Eq. (\ref{Alteqn3}) and Eq. 
(\ref{mueq3}) then follows that
\beq
\label{combmu}
\mu \dot{x} = 0.
\eeq
From the equations (\ref{combmu} and (\ref{xdot}) then follows that
\beq
\label{mu-cond}
\mu u = 0.
\eeq
The condition (\ref{combmu}), or equivalently (\ref{mu-cond}),  is thus necessary for the existence of functions $x(t)$, $y(t)$ and $z(t)$, which satisfy both the set of equations (\ref{Alteqn1}) - (\ref{Alteqn3}) and the set of equations (\ref{mueq1}) - (\ref{mueq3}) together with the constraint (\ref{dal70}) under the initial conditions (\ref{dal72}) and (\ref{dal72vel}). There are three possible cases to be considered:
\beq
\label{case1}
\mu = 0, \;u = 0,
\eeq
\beq
\label{case2}
\mu = 0, \;u \neq 0,
\eeq
and
\beq
\label{case3}
\mu \neq 0, \;u = 0.
\eeq
If the conditions (\ref{case1}) are valid, then one  one finds readily that the only common solutions of
the equations  (\ref{Alteqn1}) - (\ref{Alteqn3}) and the equations (\ref{mueq1}) - (\ref{mueq3}) as well as   the constraint equation (\ref{dal70}) which satisfy the initial conditions are the following,
\beq
\label{sol1}
x = 0, \; y = 0, \; z = wt.
\eeq
Likewise, if the conditions (\ref{case2}) are in force, then the only possible solutions are
\beq
\label{sol2}
x = ut, \; y = 0, \; z = 0.
\eeq
Finally, if the conditions (\ref{case3}) are valid, one finds the following solution,
\beq
\label{sol3}
x = 0, \; y = 0, \; z = 0, \; \mu = c,
\eeq
where $c$ is a constant.

The solutions (\ref{sol1}, (\ref{sol2}), and (\ref{sol3}), respectively, are the only functions which satisfy both the Lagrange equations of motion in the form (\ref{Alteqn1})  - (\ref{Alteqn3}) and the variational equations  (\ref{mueq1}) - (\ref{mueq3}) together with the constraint (\ref{dal70}), under the initial conditions (\ref{dal72}) and (\ref{dal72vel}). These solutions are clearly exceptional in that the non-holonomic constraint (\ref{dal70}) is no constraint at all for these solutions.  

\section{Existence of Lagrangians and Hamiltonians in Pars' example}

It should be observed that the fact that the variational procedure involving the multiplication rule does not lead to equations of motion identical to those which follow from the generalized principle of d'Alembert  in the case of non-holonomic systems,  does not prove that there is no variational principle at all for non-holonomic systems.  One may still wonder whether non-holonomic systems may nevertheless admit some kind of variational formulation. A straightforward answer to this question is obtained if one can show that the correct equations of motion (\ref{Lagnh3}) together with the constraints (\ref{nholco}) constitute a set of Lagrangian equations with some appropriate Lagrangian. This is an inverse problem, which is trivial in the case of one-dimensional systems. Complete results on the inverse problem in  question exist  for two-dimensional systems, but not for systems of dimension three or higher.  We analyze the problem posed here only in the  non-holonomic three-dimensional special case considered  by Pars, which was analyzed in some detail above.

The question is now whether the equations (\ref{Alteqn1}) - (\ref{Alteqn3}) are the Euler-Lagrange equations  with some appropriately chosen Lagrangian, or linearly equivalent to such Euler-Lagrange equations in three space dimensions. For this problem we refer to a paper by Douglas  \cite{Douglas}  on the inverse problem in the calculus of variations as well as to a paper by Crampin {\em et al.} \cite{Crampin},  which gives a geometric formulation of the inverse problem, with due reference to the paper of Douglas.

Using results given in the papers by Douglas and Crampin {\em et al.}, referred to above, one finds that the equations  (\ref{Alteqn1}) - (\ref{Alteqn3}) can \underline{not}  be recast into  linearly equivalent  equations involving three variables, such that these equivalent equations  are the Euler-Lagrange equations of some appropriate functional.  We know that the space of dynamically accessible paths in the problem under consideration is in fact two-dimensional, so it is then natural to look for a variational formulation in a two-dimensional space. It will be shown that the system of equations (\ref{Alteqn1}) - (\ref{Alteqn3}) can be reduced to an equivalent  two-dimensional autonomous system,  for which there exist Lagrangians. 

Eliminating the quantity $\dot{x}$ from  equation in the system (\ref{Alteqn2}) with the aid of the relation 
(\ref{xdot}) above, one obtains a two-dimensional autonomous system from the equations (\ref{Alteqn1}) - (\ref{Alteqn3}) , which involves the variables $y$ and $z$ only,
\bea
\label{2dimeq}
\ddot{y} & = &\ell^{2} u\dot{z}\left (\ell^{2} + z^{2} \right )^{-\frac{3}{2}}, \\
\ddot{z} & = & 0.\nonumber
\eea
The simple system of equations (\ref{2dimeq}) is indeed obtainable from a principle of stationary action in a space of two dimensions. There is in fact more than one Lagrangian for which the equations (\ref{2dimeq}) are the Euler-Lagrange equations. It is known that Lagrangians  which are derived from the equations of motion are not necessarily unique. \cite{Okubo}  We display below two such Lagrangians $L_{I}$ and  $L_{II}$, whose difference is not a time derivative of some appropriate function:
\beq
\label{ell1}
L_{I} := m\dot{y}\dot{z} - \frac{muz\dot{z}}{\sqrt{\ell^{2}+z^{2}}} \log \left ( \frac{\dot{z}}{c_{0}}\right ),
\eeq
where $c_{0}$ is a constant with the dimension of velocity. It should  be noted that the second equation in (\ref{2dimeq}) implies that $\dot{z}(t)$ is of constant sign for  $t > 0$. The sign of the constant $c_{0}$
in Eq. (\ref{ell1}) should  be chosen to be the same as the sign of the initial value $w$, so that 
$\dot{z}(t)/c_{0} > 0$ for $t>0$. It should also be noted that the absolute value of the dimensional constant $c_{0}$ is of no consequence for the equations of motion. The difference of two Lagrangians corresponding to two different constants  $c_{0}$ and $c_{0}'$, respectively, is 
\beq
\label{diffc} 
mu \left( \log\frac{c_{0}}{c_{0}'} \right ) \frac{d}{dt} \sqrt{\ell^{2} + z^{2}}.
\eeq
Since the difference (\ref{diffc}) is a time derivative, the  Lagrangians corresponding to the different constants $c_{0}$ and $c_{0}'$ are equivalent. 

The second Lagrangian is,
\beq
\label{ell2}
L_{II} := \frac{m \dot{y}\dot{z}^{2}}{2c_{0}} -  \frac{muz\dot{z}^{2}}{c_{0}\sqrt{\ell^{2}+z^{2}}}, 
\eeq
where $c_{0}$ is again a constant with the dimension of velocity. 

In the Lagrangians above, the mass $m$ occurs only as a multiplicative factor.  It has been included for convenience in order  to keep track of the dimensions in the Lagrangians in question.

The construction of the Lagrangians (\ref{ell1}) and (\ref{ell2}) using the methods developed in Douglas paper \cite{Douglas}  is a  task which involves lengthy calculations, which we will not present here. However one can easily verify that the curious  Lagrangians $L_{I}$ and $L_{II}$  above  do give rise to  the system of equations (\ref{2dimeq}). We will consider the Lagrangian $L_{I}$ above only, leaving the detailed consideration of the second Lagrangian $L_{II}$ to the interested reader.

From Eq. (\ref{ell1}) follows readily that
\beq
\label{dyell1}
\frac{\partial L_{I}}{\partial y} = 0,
\eeq
and 
\beq
\label{ddotyell1}
\frac{\partial L_{I}}{\partial \dot{y}} = m \dot{z},
\eeq
whence the Euler equation
\beq
\label{Euly}
\frac{\partial L_{I}}{\partial y} - \frac{d}{dt}\left (\frac{\partial L_{I}}{\partial \dot{y}}\right ) = 0,
\eeq
implies  that
\beq
\label{firstLag}
\ddot{z} = 0.
\eeq
Likewise,
\beq
\label{dzell1} 
\frac{\partial L_{I}}{\partial z} = -\frac{m\ell^{2}u\dot{z}}{(\ell^{2}+z^{2})^{\frac{3}{2}}}\left ( \log \frac{\dot{z}}{c_{0}} \right ),
\eeq
and 
\beq
\label{ddotzell1}
\frac{\partial L_{I}}{\partial \dot{z}} = m\dot{y}  - \frac{muz}{\sqrt{\ell^{2}+z^{2}}}\left [1 + \log \left (\frac{\dot{z}}{c_{0}} \right ) \right ].
\eeq
Inserting the expressions (\ref{dzell1}) and (\ref{ddotzell1}) in the appropriate Euler equation
\beq
\label{Eulz}
\frac{\partial L_{I}}{\partial z} - \frac{d}{dt}\left (\frac{\partial L_{I}}{\partial \dot{z}}\right ) = 0,
\eeq
one obtains the expression
\beq
\label{secLag}
\ddot{y} = \frac{\ell^{2}u\dot{z}}{(\ell^{2}+z^{2})^{\frac{3}{2}}}+ \frac{uz}{(\ell^{2}+z^{2})^{\frac{1}{2}}} \frac{\ddot{z}}{\dot{z}}.
\eeq
Using the equation (\ref{firstLag}) in the expression (\ref{secLag}) above, one ends up with
the equation
\beq
\label{secLagr2v}
\ddot{y} = \ell^{2}u\dot{z}(\ell^{2}+z^{2})^{-\frac{3}{2}}.
\eeq
This demonstrates that the equations of motion (\ref{2dimeq})  are equivalent to the Euler equations with the Lagrangian $L_{I}$ given in  Eq. (\ref{ell1}).

For completeness we also record the Hamiltonian $H_{I}$ which corresponds  to the Lagrangian
 (\ref{ell1}).  The canonical momenta are defined in the standard manner,
 \beq
 \label{canmomy}
 p_{y} := \frac{\partial L_{I}}{\partial \dot{y}} = m\dot{z},
 \eeq
 and
 \beq
 \label{canmomz}
 p_{z}:= \frac{\partial L_{I}}{\partial \dot{z}} = m\dot{y} -  \frac{muz}{\sqrt{\ell^{2}+z^{2}}}\left [1 + \log \left  ( \frac{\dot{z}}{c_{0}} \right ) \right ].
 \eeq
 The equations (\ref{canmomy}) and (\ref{canmomz}) can be solved for the  velocities $(\dot{y}, \dot{z})$ in terms of the canonical momenta $(p_{y}, p_{z})$,
 \beq
\label{veloz}
\dot{z} = \frac{p_{y}}{m},
\eeq
and
 \beq
 \label{veloy}
 \dot{y} = \frac{p_{z}}{m} +  \frac{uz}{\sqrt{\ell^{2}+z^{2}}}\left [1 +\log  \left (\frac{p_{y}}{mc_{0}} \right ) \right ].
 \eeq
This  leads to the following Hamiltonian $H_{I}$,
 \beq
 \label{Ham1}
 H_{I} = \frac{1}{m}p_{y}p_{z} +  \frac{uz}{\sqrt{\ell^{2}+z^{2}}}p_{y}\log \left (\frac{p_{y}}{mc_{0}}\right ).
\eeq
The Hamiltonian equations involving the Hamiltonian $H_{I}$ in Eq. (\ref{Ham1}) are as follows,
\beq
\label{Ha}
\dot{y} :=  \frac{\partial H_{I}}{\partial p_{y}} = \frac{1}{m} p_{z} + \frac{ uz}{\sqrt {\ell^{2} + z^{2}}}[1+ \log \left (\frac{p_{y}}{mc_{0}} \right )],
\eeq
\beq
\label{Hb}
\dot{z} :=  \frac{\partial H_{I}}{\partial p_{z}} = \frac{1}{m}p_{y},
\eeq
\beq
\label{Hc}
\dot{p}_{y} := -  \frac{\partial H_{I}}{\partial y} = 0,
\eeq
and
\beq
\label{Hd}
\dot{p}_{z} :=  - \frac{\partial H_{I}}{\partial z} =  - \frac{\ell^{2} u}{(\ell^{2}+z^{2})^{\frac{3}{2}}} p_{y}\log \left (\frac{p_{y}}{mc_{0}}\right ).
\eeq
Combining Eqns. (\ref{Ha}) and (\ref{Hd}) one obtains the first equation in the two-dimensional system (\ref{2dimeq}). Likewise, combining Eqns. (\ref{Hb}) and (\ref{Hc}) one obtains the second equation in (\ref{2dimeq}).

Using the Lagrangian $L_{II}$ given above as a starting point, one obtains an alternative canonical Hamiltonian formulation of the two-dimensional system of equations (\ref{2dimeq}).  The canonical momenta $p_{y}$ and $p_{z}$ are now related to the velocities $\dot{y}$ and $\dot{z}$ as follows,
\beq
\label{mom2y}
p_{y} = \frac{m\dot{z}^{2}}{2c_{0}}, 
\eeq
and
\beq
\label{mom2z}
p_{z} =  \sqrt{\frac{2mp_{y}}{c_{0}}}\left [ \dot{y} - \frac{2uz}{\sqrt{\ell^{2}+z^{2}}} \right ].
\eeq
The Hamiltonian $H_{II}$ is 
\beq
\label{Ham2}
H_{II} = \sqrt{\frac{2c_{0}p_{y}}{m}} p_{z} + \frac{2uz}{\sqrt{\ell^{2}+z^{2}}}p_{y}.
\eeq
It is readily verified that the canonical equations with the Hamiltonian (\ref{Ham2}) likewise reproduce the equations (\ref{2dimeq}).

\section{Concluding remarks}

We consider the  principle of stationary action   for  autonomous mechanical systems. The principle of stationary action, Hamilton's principle, can be generalized to  systems with {\em holonomic} constraints by including the constraints in the variational procedure by means of the multiplication rule in the calculus of variations. The  corresponding procedure in the case of {\em non-holonomic} constraints leads to equations which are not, in general,  identical  to the correct equations of motion. This fact has  been known  for some  time.

It has been shown,  in the case of a  {\em particular} three-dimensional autonomous non-holonomic system, that a variational  principle  exists; the equations of motion for the system can be reduced to a two-dimensional system of autonomous equations  which admit a variational formulation and also a canonical  Hamiltonian formulation. Explicit expressions for two nonequivalent Lagrangians  together with the corresponding Hamiltonians are given.  The example shows that one can not {\em a priori} rule out the existence of  variational principles and Hamiltonians for non-holonomic systems, however artificial.  \\


\end{document}